# Impact of Cr doping on the structure, optical and magnetic properties of nanocrystalline ZnO particles

H. S. Lokesha[1,*], P. Mohanty[1], A. R. E. Prinsloo[1] and C. J. Sheppard[1]

[1]*Cr Research Group, Department of Physics, University of Johannesburg, PO Box 524, Johannesburg, South Africa*

*Corresponding author email: lokeshahs@uj.ac.za

**Abstract**

The role of Cr incorporation into the ZnO were probed through investigations into the structural, optical and magnetic properties. $Zn_{1-x}Cr_xO$ with $x$ = 0, 0.01, 0.03 and 0.05, nanoparticles were prepared by solution combustion method. Powder x-ray diffraction (XRD) results reveal, all the synthesized samples are in single hexagonal wurtzite crystal structures, indicating that $Cr^{3+}$ ions substitute the $Zn^{2+}$ ions without altering the structure. The crystallite size and microstrain were calculated using the Willamson−Hall method and found to be $36 \pm 2$ nm for ZnO and it reduced with the increase of Cr dopant concentration to $20 \pm 2$ nm for $Zn_{0.95}Cr_{0.05}O$. Rietveld refinement analysis revealed that the lattice parameters $a$ and $c$ of ZnO are well matched with standard data (JCPDS #36-1451). The value of both $a$ and $c$ increases slightly with the increase in Cr concentration. Transmission electron microscopy (TEM) revealed that the particle size were $48 \pm 2$ nm, $29 \pm 2$ nm and $25 \pm 2$ nm for the $Zn_{1-x}Cr_xO$ with $x$ = 0, 0.03 and 0.05, respectively. TEM morphology indicated particles are agglomerated in the doped samples. The band-gap decreases slightly from $3.305 \pm 0.003$ eV to $3.292 \pm 0.003$ eV with increase of Cr content from $x$ = 0 to 0.05, respectively. Photoluminescence measurements revealed the presence of defects in the samples, associated with zinc vacancies ($V_{Zn}$) and singly ionized oxygen vacancy ($V_o^+$). The field-dependent magnetization $M(\mu_oH)$ measurements of ZnO and Cr-doped ZnO were carried out using a vibrating sample magnetometer (VSM) at 300 K. All the samples exhibits ferromagnetic behavior. This long-range ferromagnetism ordering observed in ZnO is explained based on bound magnetic polaron (BMP) mechanism. The $V_o^+$ vacancies playing a crucial role in observed room temperature ferromagnetism (RTFM) in ZnO. There is a sufficient amount of BMPs formed in Cr doped ZnO because of the defects ($V_{Zn}$) present in these samples. Therefore, the overlapping of BMPs results in the RTFM. However, the antiferromagnetic coupling at a higher doping concentration of Cr, weakens the observed RTFM.

## 1. Introduction

In order for diluted magnetic semiconductor (DMS) materials to find practical applications in spintronic devices, it is necessary to obtain both the Curie temperature ($T_C$) and ferromagnetic properties at or above room temperature [1]. The origin of the ferromagnetism (FM) in the DMS materials is attributed to intrinsic defects, and not from the presence of magnetic clusters of dopants [2]. The formation of the intrinsic defects in the undoped material is mainly dependent on the synthesis procedure [3], while the dopant transition metal (TM) will also create defects [2,3].

ZnO has been considered as one of the most promising host materials for fabricating DMS because of its unique properties, including a wide band-gap (3.3 eV), high solubility for TM doping and large exciton binding energy of 60 meV [4]. Room temperature ferromagnetism (RTFM) is also reported in undoped ZnO nanoparticles and is attributed to the presence of the point defects, which are strongly dependent on synthesis methods [5–7]. Gao et al. [5] reported that in a ZnO sample annealed in a mixed atmosphere of nitrogen and oxygen, the RTFM decreases after annealing in a rich-oxygen atmosphere. Thus, the oxygen vacancies plays a significant role for in the origin of RTFM in ZnO [5]. The origin of RTFM in undoped ZnO has been ascribed to singly occupied oxygen vacancy mediated oxygen vacancy clusters [6], singly occupied oxygen vacancy ($F^+$ center) [7], Zn interstitial [8] and Zn vacancy [9].

The TM doped ZnO became one of the promising DMS materials because of the existence of RTFM [3,4,10–17]. Cr, which is an antiferromagnet (AFM) at room temperature [18], can effectively be doped at the $Zn^{2+}$ positions, since the ionic size of $Cr^{3+}$ (0.61 Å) is smaller than that of $Zn^{2+}$ (0.74 Å) [19–21]. Duan et al. [22] reported that the RTFM in Cr-doped ZnO attributed to super-exchange Cr–O–Cr interactions and the dopant and donor defects (zinc interstitial, oxygen vacancies) hybridization. Photoluminescence (PL) of ZnO and Cr-doped ZnO shows a peak around 410 nm that is correlated to interstitial Zn, as well as Zn vacancies [23]. The peak between 530 and 640 nm is correlated to oxygen defects with different charge states, including natural, singly and doubly occupied oxygen vacancies [4,24,25]. Liu et al. [21] reported that the exchange interactions between Cr 3d and O 2p spin moments are responsible for RTFM in Cr-doped ZnO nanoparticles. However, Liu et al. [24] concluded that

Zn interstitials and Cr doping account for RTFM in Cr-doped ZnO powder using PL measurements and the BMP model.

In this paper, focusses the impact of Cr-doping on the structure, optical and magnetic properties of ZnO. The samples were synthesized by solution combustion method. The structure and morphology of powdered samples were analyzed by XRD and HR-TEM. The band-gap and nature of defects/vacancies present in the samples were analyzed using diffuse reflectance spectroscopy (DRS) and PL studies, respectively. The RTFM was measured using vibrating sample magnetometer (VSM). From last decade, the origin of room temperature ferromagnetism (RTFM) in undoped and TM doped ZnO was highly debated because it was unclear whether the RTFM is because of intrinsic or magnetic ion clusters [26–28]. In the present study a detailed experimental investigation is reported in order to identify the origin of RTFM in undoped and Cr-doped ZnO.

1. **Experiment details**

*2.1 Material synthesis*

$Zn_{1-x}Cr_xO$ with $x$ = 0, 0.01, 0.03, and 0.05, powder samples were prepared by solution combustion method [29]. For the samples synthesis, $Zn(NO_3)_2 \cdot 6H_2O$ (98%), $NH_2CH_2COOH$ (99%) and $Cr(NO_3)_3 \cdot 9H_2O$ (99%) were used as initial materials without further purification. The amount of the nitrates used were based on the stoichiometric condition of oxidizer to fuel ratio that is unity and dissolved in 30 ml double distilled water. The mixture was magnetically stirred until it became transparent. The solution is then placed into a pre-heated (at 350 ± 10 ℃) muffle furnace. Initially, the solution boils, then ignites and burns. Finally, a voluminous foamy product (ash) was obtained. The product was ground into a fine powder using an agate mortar and pestle.

*2.2 Characterization*

The phase purity of the samples was analyzed through x-ray diffraction techniques, utilizing a Phillips PAN analytical X-pert Pro X-ray diffractometer (Cu–$K\alpha$ with $\lambda$ = 1.54056 Å). XRD patterns of the samples were recorded in a $2\theta$ range from 20 to 80°. The morphology of the samples was investigated using a transmission electron microscope (TEM) (Model: JEOL JEM−2100). In order to investigate the elemental composition of the samples using energy dispersive x-ray spectroscopy (EDS), the detector of Oxford Instruments attached to the TEM was used. The diffuse reflectance spectra (DRS) were recorded using a laboratory spectrometer instrument [30]. PL spectra were recorded using an excitation wavelength of 248 nm at 300 K. The magnetic measurement was carried out using a vibrating sample magnetometer (Mode: Lake Shore−7410 series) at room temperature.

# 3  Results and discussion

## 3.1  X-ray diffraction

Figure 1 shows the Rietveld refinement of XRD patterns of as prepared $Zn_{1-x}Cr_xO$, with $x = 0$, 0.01, 0.03, and 0.05. All the diffraction peaks observed within the measurement of $2\theta$ range (20−80º) are characteristic of hexagonal wurtzite structure of ZnO with space group P63mc (PDF#36-1451). The broadening (considering the full width at half maximum) of the XRD peaks increased with an increase in the Cr ion doping concentration, as can be seen in the main peaks associated with the (100), (002) and (101) planes shown in figure 2(a), suggesting a decrease of crystallite size or change of lattice strain or combination of both [31]. A negligible peak shift is observed after doping with a small Cr concentration ($\leq 5\%$) that can be ascribed to the $Cr^{3+}$ (0.61 Å) replacing $Zn^{2+}$ (0.74 Å) sites [32]. In addition, the intensity of the XRD peaks progressively decreases with Cr doping concentration. This can be attributed to the change in atomic scattering factor [33] or loss of crystallinity because of lattice distortion [34]. In previous reports [22,35] the Cr-doped ZnO was synthesized using the combustion method with glycine and these showed a secondary zinc chromate phase when Cr concentrations exceed 0.05 mol%. In the present work, no secondary phases, including binary zinc chromium phases, are identified within the detection limit of the XRD. The average crystallite size ($D$) and microstrain were determined using Williamson−Hall (W−H) relation [36]. The details of the method were reported elsewhere [37]. The plot of $(\beta \cos\theta/\lambda)$ as a function of $(4\sin\theta/\lambda)$ is expected to be linear, where $\beta$ is the full width half maximum and $\lambda$ is the wavelength of x-ray. The W−H plots of $Zn_{1-x}Cr_xO$, with $x = 0$, 0.01, 0.03, and 0.05 are shown in figure 2 (b). A linear fit of the straight line provides the information of average crystallite size (inverse of intercept of the straight line) and microstrain (slope of the line) of the sample. The calculated structural parameters of all samples are tabulated in Table 1. The crystallite size decreases and the microstrain increases with the increase of Cr ion concentration up to $x = 0.05$. The decrease in crystallite size is because of the doping of Cr into the ZnO lattice prevents crystallite growth, possibly slows down the formation of grain boundaries and/or modify the rate of nucleation during the sample crystallization [32,34]. Doping of Cr into Zn site could alter the geometrical structure of the ZnO because of the difference in the ionic radii between Cr and Zn, thus, the lattice strain is expected to increase with the increase of Cr content.

The unit cell parameters were calculated as a function of Cr ion concentration through the Rietveld refinement of the XRD data using GSAS II software [38]. The refined parameters such as lattice parameters, cell volume, atom position and fitting parameters are tabulated in Table 1. The cell parameters values are comparable with standard data (JCPDS#36-1451). The lattice parameters slightly increase with increase in the Cr concentration up to $x = 0.05$. This is related to the difference between the ionic radii of the cations, leading to the lattice distortion and the strain-induced during the combustion synthesis [39,40].

*3.2 Transmission electron microscope*

Figure 3 (a–c) shows the TEM micrographs for the $Zn_{1-x}Cr_xO$, with $x = 0$, 0.03 and 0.05, respectively. The HR-TEM image and particles distribution of the corresponding samples are depicted in figure 3 (d–i). It is found that, the incorporation of Cr ions into ZnO strongly influenced on the morphology of the samples, i.e., the particle size decreased while the particles' agglomeration increased with Cr concentration. A few particles appear much larger in size in the Cr-doped samples, but this is because of the agglomeration of smaller particles and growth during the combustion process. Particles are nearly spherical in shape and but non-uniform in size, as is seen in the histograms shown in figure 3 (g–i). The particle size distributions of the samples were fitted to a log-normal function and the average particle sizes were determined to be 48 ± 2 nm, 29 ± 2 nm and 25 ± 2 nm for the $Zn_{1-x}Cr_xO$, with $x = 0$, 0.03 and 0.05, respectively. Furthermore, the spacing between two adjacent fringes was calculated from HR-TEM images and found to be 0.278 nm for ZnO. The measured *d*-spacing values were compared with the standard data observed in the PDF#36-1451 and is in agreement with the value of 0.281 nm which is assigned to (100) plane. In the case of $Zn_{1-x}Cr_xO$, with $x = 0.03$ and 0.05, the *d*-spacing value found to be 0.52 nm, which is comparable with lattice constant ($c = 0.5213$ nm) of ZnO grown along the *c*-axis direction [5]. Energy dispersive x-ray spectroscopy (EDX) results provide information about the chemical compositions in the samples, there is no additional elements in the samples.

*3.3 Diffuse Reflectance Spectroscopy*

In order to examine the band-gap of the powder samples, the DRS spectra were used. DRS spectra of $Zn_{1-x}Cr_xO$, with $x = 0$, 0.01, 0.03 and 0.05, were measured in the spectral range 250 to 750 nm and are shown in figure 4. It is observed that, the reflectance in the wavelength range of 400 to 700 nm decreases with increased concentration. According to the Kubelka−Munk theory [41], the graph plotting $[F(R)h\nu]^2$ as function of photon energy ($h\nu$) for $Zn_{1-x}Cr_xO$, with $x = 0$, 0.01, 0.03 and 0.05, are shown as in figure 5. Here $F(R) = \frac{(1-R)^2}{2R}$ is the Kubelka–Munk function and $R$ is the reflectance. The band-gap of the samples was determined by extrapolating the linear portions of the plots onto the $x$–axis ($h\nu$). The band-gap values of the samples are tabulated in Table 1. The band-gap value decreases with an increase in Cr concentration, this is attributed to $sp$–$d$ exchange interaction between the band electrons and localized $d$ electrons of the Cr ions [42]. The band-gap narrowing as compared to that of ZnO with the addition of Cr, is linked to $s$–$d$ and $p$–$d$ exchange interactions causing a negative and a positive correction in the conduction band and the valence-band, respectively [43].

*3.4 Photoluminescence*

Figure 6 shows PL emission spectra of $Zn_{1-x}Cr_xO$ ($x = 0$, 0.01, 0.03, and 0.05), recorded under excitation wavelength 248 nm at room temperature. All the emission spectra consist of two peaks, a sharp peak is observed around 384 nm and a broad emission peak is seen from 400 to 625 nm. The peak at 384 nm is due to near-band edge emission and originates from the radiative recombination of free excitons [32]. PL peak intensities decrease dramatically with an increase of Cr concentration, signifying a quenching of PL emissions [42]. The broad emission peak is observed because of the presence of different recombination sites and defects [32]. Liu et al. [24] reported that the PL of Cr-doped ZnO, the spectra show peaks at 405, 418, 438, 466 and 490 nm assigned to zinc vacancy ($V_{Zn}$), oxygen interstitials ($O_i$), zinc interstitials ($Zn_i$), singly negatively charged Zn vacancy ($V_{Zn}^-$) and oxygen vacancy ($V_o$), respectively. Punia et al. [44] reported on the oxygen vacancies present in three different charge states in Li-doped ZnO, PL peaks observed around 530, 575 and 645 nm, corresponding to neutral oxygen vacancies ($V_o$), single charge oxygen vacancies ($V_o^+$) and double charge oxygen vacancies ($V_o^{++}$), respectively [44]. The presence of oxygen vacancies in Cr doped ZnO was reported from the analysis of x-ray photoelectron spectroscopy [45] and extended x-ray absorption fine structure [32]. Further, in order to recognize the type of vacancy that contributes to the broad emission in the visible region, the broad peak (425−625 nm) de-convoluted into three peaks using Gaussian fits centered at 484, 548 and 605 nm for ZnO, as shown in figure 7. These peaks are shifted towards the blue region, with the peaks centered at 481, 530 and 592 nm, in Cr-doped ZnO shows reduced peak width. Also, a small intense peak is observed in all the samples around 410 nm, which finds its origin in Zn vacancy [23]. The blue-green band (481 nm) is attributed to electron transition from donor level of zinc interstitial ($Zn_i$) to acceptor level of zinc vacancy ($V_{Zn}$) near valence band [44,45] and surface defects [32]. This peak intensity decreased with increase of Cr concentration. The peak around 530 and 600 nm corresponds to singly ionized oxygen vacancies ($V_o^+$) [4,32,48–50] and doubly ionized oxygen vacancies ($V_o^{++}$) [48,50,51], respectively. In several reports [4,50,52] the emission peak around 600 nm is related to oxygen interstitial ($O_i$) defects. The visible PL emissions observed are produced from intrinsic defects and not from any impurity phase as is evidenced by XRD.

*3.5 Magnetic properties*

In figure 8, the magnetization as a function of applied magnetic field $M(\mu_o H)$ curves of the $Zn_{1-x}Cr_xO$ ($x$ = 0, 0.03 and 0.05) samples was measured at room temperature. The hysteresis loops signify that the $Zn_{1-x}Cr_xO$ ($x$ = 0, 0.03 and 0.05) samples show RTFM, as is indicated by coercive fields of 228, 190 and 224 Oe, respectively. The saturation magnetization ($M_s$) is found to be 0.867, 0.674 and 0.627 emu.g$^{-1}$ for the $Zn_{1-x}Cr_xO$ samples with $x$ = 0, 0.03 and 0.05, respectively. As compared to literature, the combustion synthesized $Zn_{1-x}Cr_xO$ ($x$ = 0.03 and 0.05) shows higher $M_s$ values when compared to samples prepared using sol-gel (2.2×10$^{-3}$ emu.g$^{-1}$) [21] and hydrothermal methods (2×10$^{-2}$ emu.g$^{-1}$) [53]. The variation in $M_s$ depends on many factors that affect the magnetic behavior, including the creation of strain, vacancies or defect states and particle size [54]. The origin of RTFM in ZnO is considered to be intrinsic because the presence of impurities in ZnO is ruled out by the XRD and EDX analysis. The visible PL emission showed the presence of intrinsic defects states in ZnO ($V_{Zn}$, $V_o^+$ and $V_o^{++}$). The neutral oxygen vacancies ($V_o$) and /or $V_o^{++}$ vacancies have spin-zero ground states, so these do not contribute to the FM in ZnO [7]. The $V_o^+$ vacancies can activate bound magnetic polarons (BMP) in ZnO DMS [55]. The number of BMPs concentration is found to be 5×10$^{18}$ cm$^{-3}$, this represents the direct interaction of the BMPs leading to ferromagnetic domains in ZnO. However, the origin of long-range ferromagnetic ordering in pure ZnO does not exclude other types of defects such as Zn vacancies [55].

In the case of Cr doped ZnO, the highest $M_s$ is observed in $Zn_{0.97}Cr_{0.03}O$ and found to be 0.675 emu.g$^{-1}$ and the $M_s$ value decreased to 0.626 emu.g$^{-1}$ for $Zn_{0.95}Cr_{0.05}O$. This can be explained as follows; with an increase in Cr concentration, the Cr atoms occupy adjacent cation sites in ZnO lattice resulting in antiferromagnetically coupled spin interactions, leading to a reduction in the $M_s$ values in $Zn_{0.95}Cr_{0.05}O$.

Furthermore, the $M(\mu_o H)$ curve is fitted with the BMP model using the following equation [16]:

$$M = M_0 L(x) + \chi_m H,$$

where the first term represents the contribution of BMP and the second term represents the paramagnetic matrix contribution. Here, the total BMP magnetization is given by $M_0 = nm_s$, where $n$ is the number of BMPs and $m_s$ is the effective spontaneous moment per BMP. $L(x)$ is the Langevin function given by [16]:

$$L(x) = \coth(x) - 1/x, \text{ with } x = m_{eff}H/k_BT,$$

where $m_{eff}$ is the true spontaneous moment per BMP and $m_s \cong m_{eff}$ at higher temperatures. The experimental $M(\mu_o H)$ curves of $Zn_{1-x}Cr_xO$ ($x$ = 0.03 and 0.05) fitted with BMP model is shown in figure 9. The fitted parameters $M_0$ and $\chi_m$ are found to be 0.63 and 0.62 emu.g$^{-1}$, 3.1×10$^{-6}$ and 1.2×10$^{-6}$ cgs for $Zn_{1-x}Cr_xO$ ($x$ = 0.03 and 0.05), respectively. The spontaneous moment per BMP ($m_{eff}$) is found to be in the order of 10$^{-18}$ emu for both samples. The number of BMPs is determined to be 2.5×10$^{18}$ and 4.5×10$^{18}$ cm$^{-3}$ for $Zn_{1-x}Cr_xO$ ($x$ = 0.03 and 0.05), respectively. The defect or BMP concentration required for long-range ferromagnetism is around 2×10$^{18}$ cm$^{-3}$, using the cation density as 3.94×10$^{22}$ cm$^{-3}$ in ZnO [16,56]. The obtained values of BMP concentration for $Zn_{1-x}Cr_xO$ ($x$ = 0.03 and 0.05) is greater than the BMP percolation threshold required for long-range ferromagnetic order in wurtzite ZnO. Recently, long-range FM at room temperature was reported for Cu-doped ZnO [3], Ag-doped ZnO [57] and Mn-doped ZnO [16]. The overlapping of BMPs causes the alignment of dopant spins, resulting in the long-range ferromagnetism at room temperature [57]. At higher Cr concentration, Cr ions occupy more of the Zn sites and possibly also defects centers, which give antiferromagnetically coupled spins and hence reduces the $M_s$ in $Zn_{0.95}Cr_{0.05}O$. Finally, the overlapping of bound magnetic polarons created because of the exchange interaction between the spin of Cr ions and defects such as $V_o^+$ and /or $V_{Zn}$ vacancies are responsible for RTFM Cr-doped ZnO.

## 4  Conclusions

In conclusion, $Zn_{1-x}Cr_xO$ ($x$ = 0, 0.01, 0.03, and 0.05) with a hexagonal wurtzite structure was synthesized through the solution combustion method. The structural parameters were investigated using Rietveld refinement of the XRD results, showing that the crystallite size decreases and microstrain increases with increase of Cr concentration. Particle sizes were determined using TEM and found to be $48 \pm 2$ nm, $29 \pm 2$ nm and $25 \pm 2$ nm for the $Zn_{1-x}Cr_xO$ with $x$ = 0, 0.03 and 0.05, respectively. In addition, the particles are agglomerated in Cr doped samples. DRS measurement showed that the band-gap decreases with the increase of Cr concentration. PL measurements illustrate near band edge emission at a wavelength of 384 nm and a broad visible emission in the range from 400 to 625 nm. The visible emissions is attributed to $V_{Zn}$, $V_o^+$ and $V_o^{++}$ vacancies. Magnetization measurements on selected samples showed RTFM in the $Zn_{1-x}Cr_xO$ samples with $x$ = 0, 0.03 and 0.05. Findings of the present work clearly shows $V_o^+$ vacancies are responsible for long-range RTFM in ZnO. The observed RTFM is explained based on the BMP model. It suggested that, the calculated values of BMPs concentration is greater than the BMP percolation threshold, so the overlapping of BMPs resulting the long-range FM in Cr-doped ZnO at room temperature.

**Acknowledgements**

Financial aid from the south African National Research Foundation (Grant No's: 120856 and 88080) and the Faculty of Science, University of Johannesburg (UJ) URC and FRC is acknowledged. The use of the NEP Cryogenic Physical Properties Measurement System at UJ, obtained with the financial support from the SANRF (Grant No: 88080) and UJ, RSA, is acknowledged. The use of Spectrum Analytical Facility at UJ is acknowledged. Prof. J.R. Botha at Nelson Mandela University, RSA, is acknowledged for completing the PL measurements

**Table**

| Refined parameters | $Zn_{1-x}Cr_xO$ | | | |
|---|---|---|---|---|
| | $x = 0$ | $x = 0.01$ | $x = 0.03$ | $x = 0.05$ |
| Crystallite size (nm) from W-H method | $36 \pm 2$ | $34 \pm 2$ | $28 \pm 2$ | $20 \pm 2$ |
| Microstrain (%) | 0.016 | 0.080 | 0.099 | 0.042 |
| Particle size from TEM (nm) | $48 \pm 2$ | -- | $29 \pm 2$ | $25 \pm 2$ |
| Lattice parameters (Å) (Error: ±0.0001) | $a = b = 3.2490$ $c = 5.2072$ | 3.2494 5.2066 | 3.2535 5.2130 | 3.2649 5.2310 |
| Cell volume (Å$^3$) | $47.62 \pm 0.02$ | $47.61 \pm 0.02$ | $47.79 \pm 0.03$ | $48.29 \pm 0.05$ |
| Band-gap (eV) | | | | |
| $\chi^2$ | 1.407 | 1.205 | 1.142 | 1.160 |
| $R_{wp}$ (%) | 4.43 | 4.14 | 4.18 | 4.21 |
| $R_F$ (%) | 2.53 | 2.25 | 2.91 | 3.44 |
| Zn $(x, y, z)$ | (0.333, 0.666, 0.009) | (0.333, 0.666, 0.003) | (0.333, 0.666, -0.006) | (0.333, 0.666, -0.006) |
| O $(x, y, z)$ | (0.333, 0.666, 0.384) | (0.333, 0.666, 0.385) | (0.333, 0.666, 0.384) | (0.333, 0.666, 0.385) |
| Cr $(x, y, z)$ | -- | (0.330, 0.660, -1.077) | (0.333, 0.666, -0.785) | (0.333, 0.666, -0.785) |

Table 1. Structural parameters of estimated from Rietveld refinement of XRD using GSAS II program. The fitted parameters chi-square ($\chi^2$), Weighted Profile ($R_{wp}$) and scale F factor ($R_F$) are given in Table 1.

**Figures**

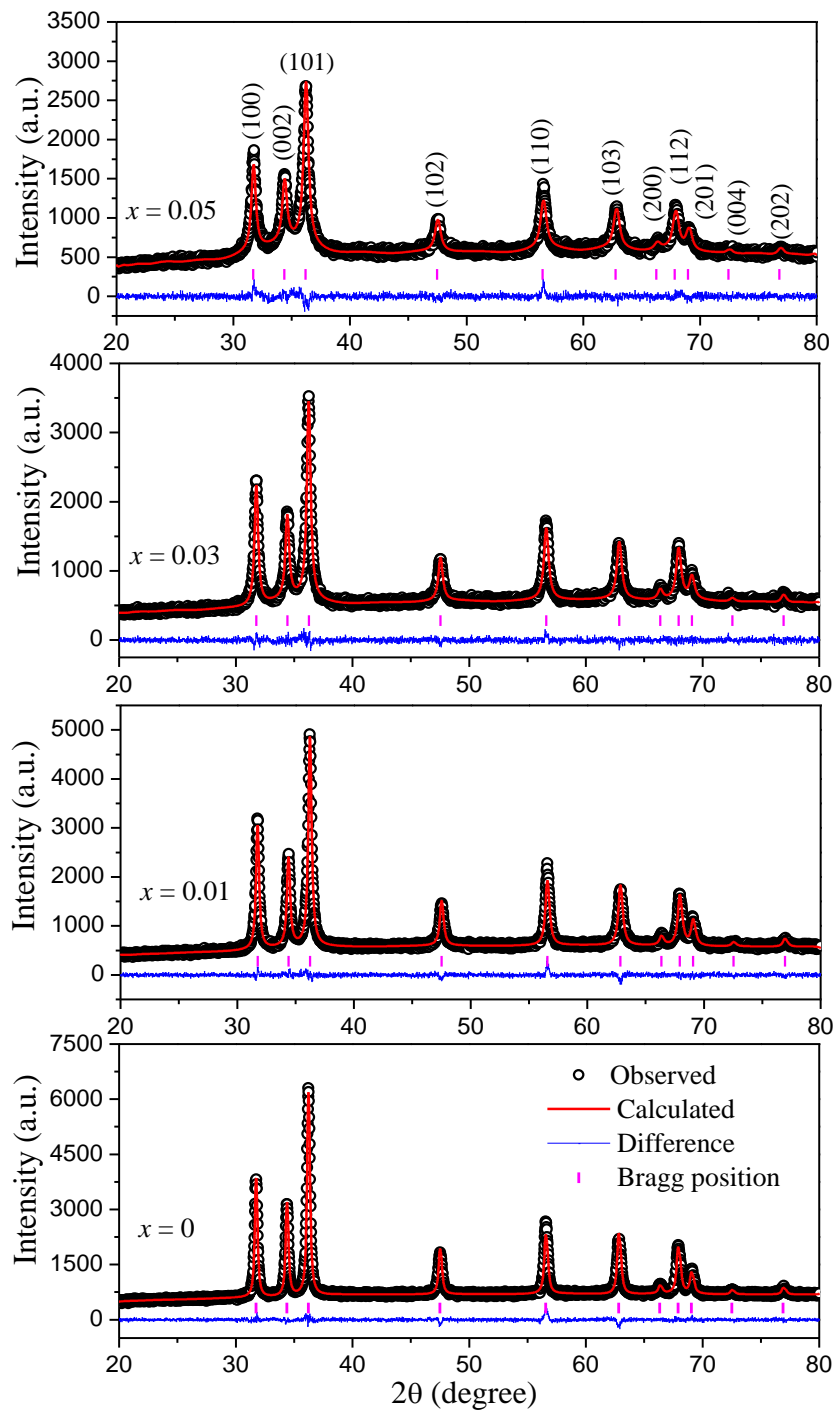

**Figure 1**. Rietveld refinement of XRD patterns of the $Zn_{1-x}Cr_xO$ ($x$ = 0, 0.01, 0.03 and 0.05) analyzed using GSAS II program. The black open circle represent the measured data compared with the calculated profile (red solid line), the small purple vertical lines below the curve are the expected Bragg positions of wurtzite structure of ZnO and the residual of the refinement shown as blue solid line.

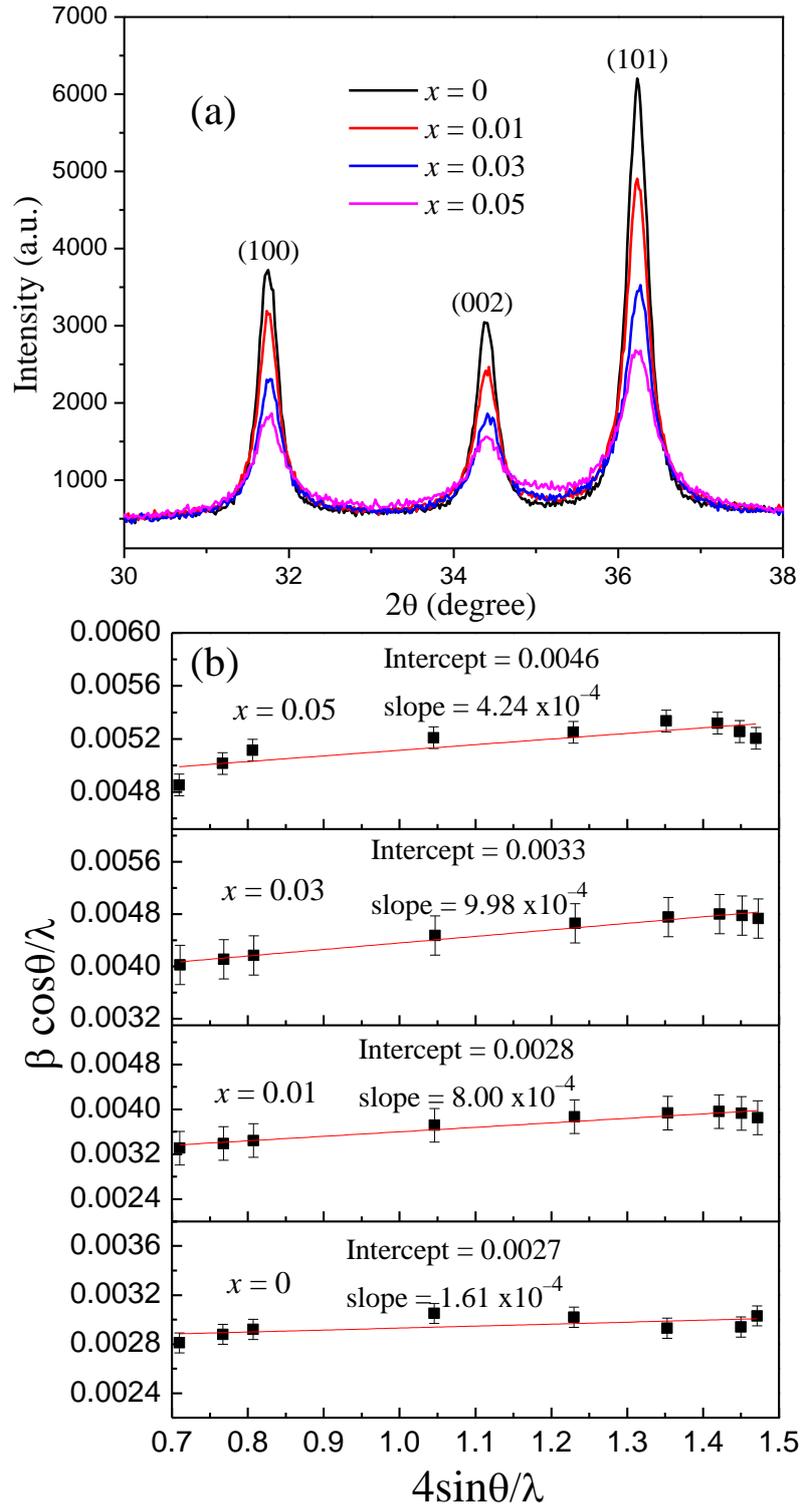

**Figure 2.** (a) The enlarged region of the (100), (002) and (101) diffraction peaks. (b) The Williamson−Hall plot, ($\beta\cos\theta/\lambda$) as function of ($4\sin\theta/\lambda$) for $Zn_{1−x}Cr_xO$ (with $x$ = 0, 0.01, 0,03 and 0.05) samples, where the red line represents linear fit of the data.

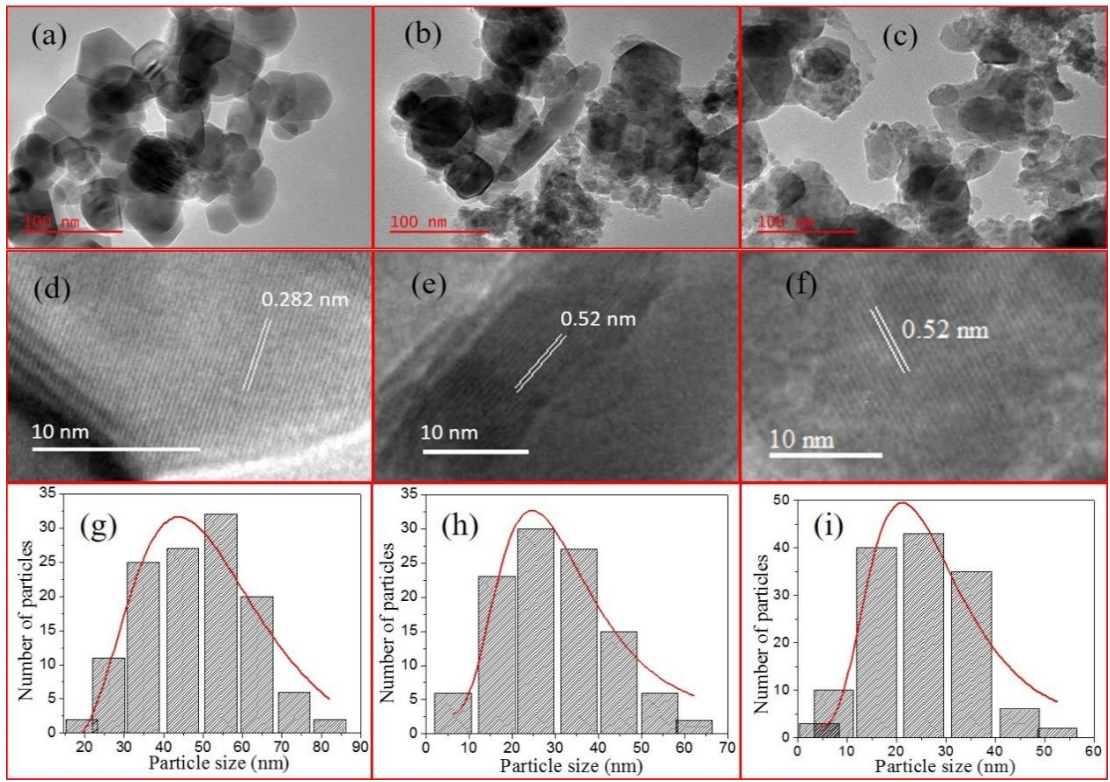

**Figure 3.** Transmission electron microscopy (TEM) images of (a) ZnO, (b) $Zn_{0.97}Cr_{0.03}O$ and (c) $Zn_{0.95}Cr_{0.05}O$ samples and their corresponding high resolution TEM images shown figure 3 (d-f). The particle size distribution of the corresponding samples are given in figure 3 (g-i), respectively.

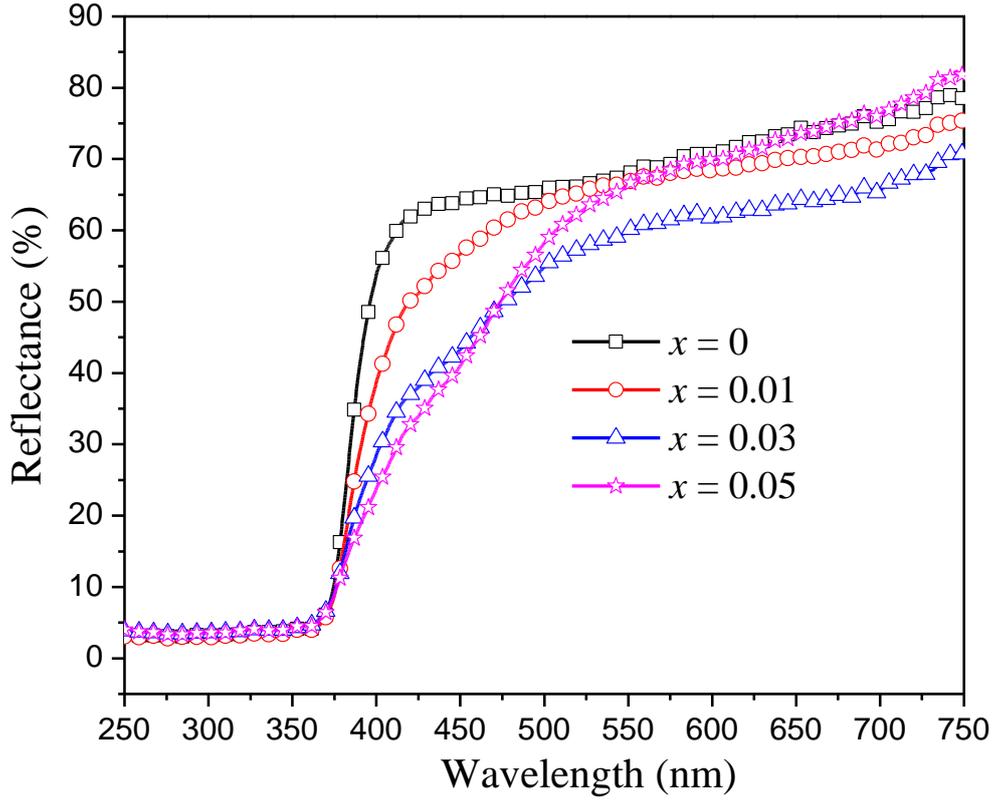

**Figure 4.** DRS spectra of $Zn_{1-x}Cr_xO$ (with $x$ = 0, 0.01, 0.03 and 0.05) samples.

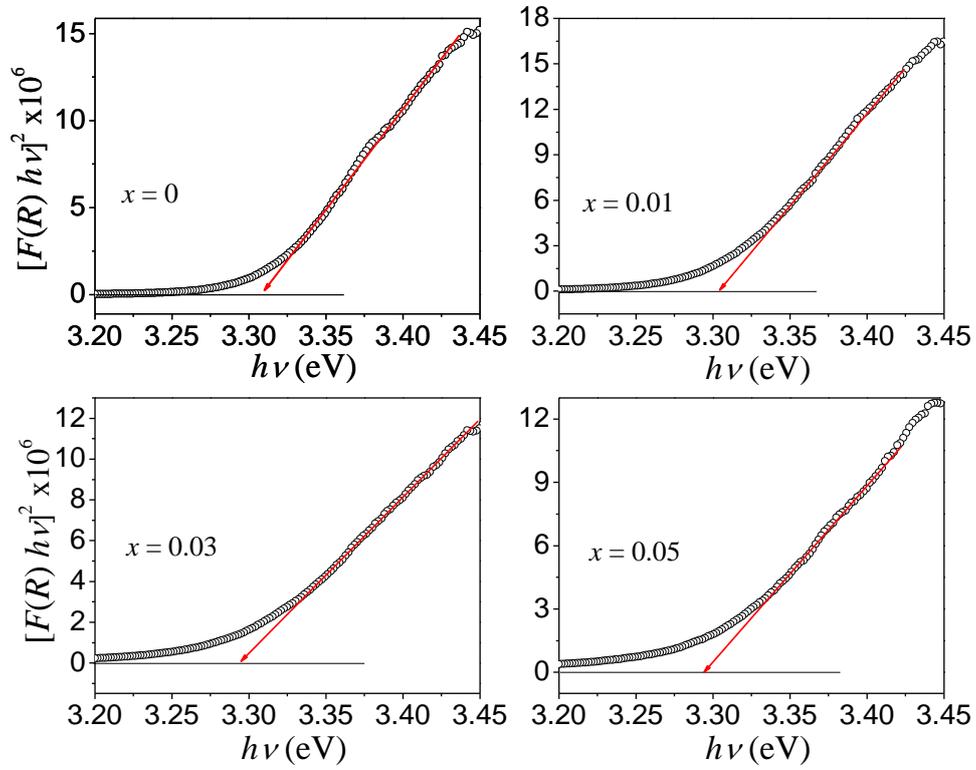

**Figure 5**. The plot of $[F(R)h\nu]^2$ as function of $h\nu$ for $Zn_{1-x}Cr_xO$ (with $x$ = 0, 0.01, 0.03 and 0.05) samples, the red line represent extended line of linear region.

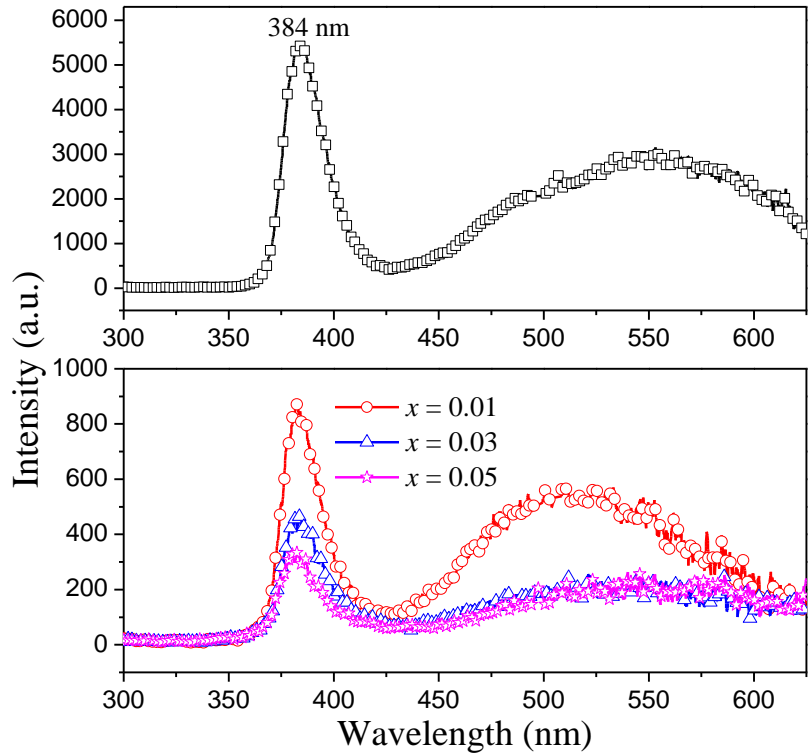

**Figure 6.** Photoluminescence spectra, showing intensity versus wavelength, of $Zn_{1-x}Cr_xO$ (with $x = 0$, 0.01, 0.03, and 0.05) measured at room temperature under excitation wavelength 248 nm.

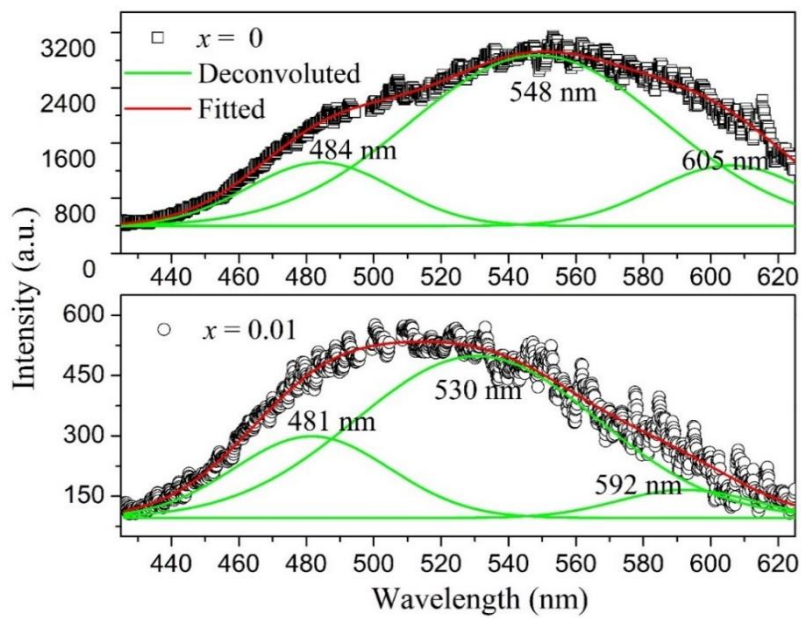

**Figure 7.** The broad visible emission spectra of $Zn_{1-x}Cr_xO$ ($x = 0$ and 0.01) fitted into three peaks with Gaussian functions. The symbol represents experimental data, red (thick) solid line is a fit of experimental data with Gaussian function and green (thin) solid lines are deconvoluted peaks.

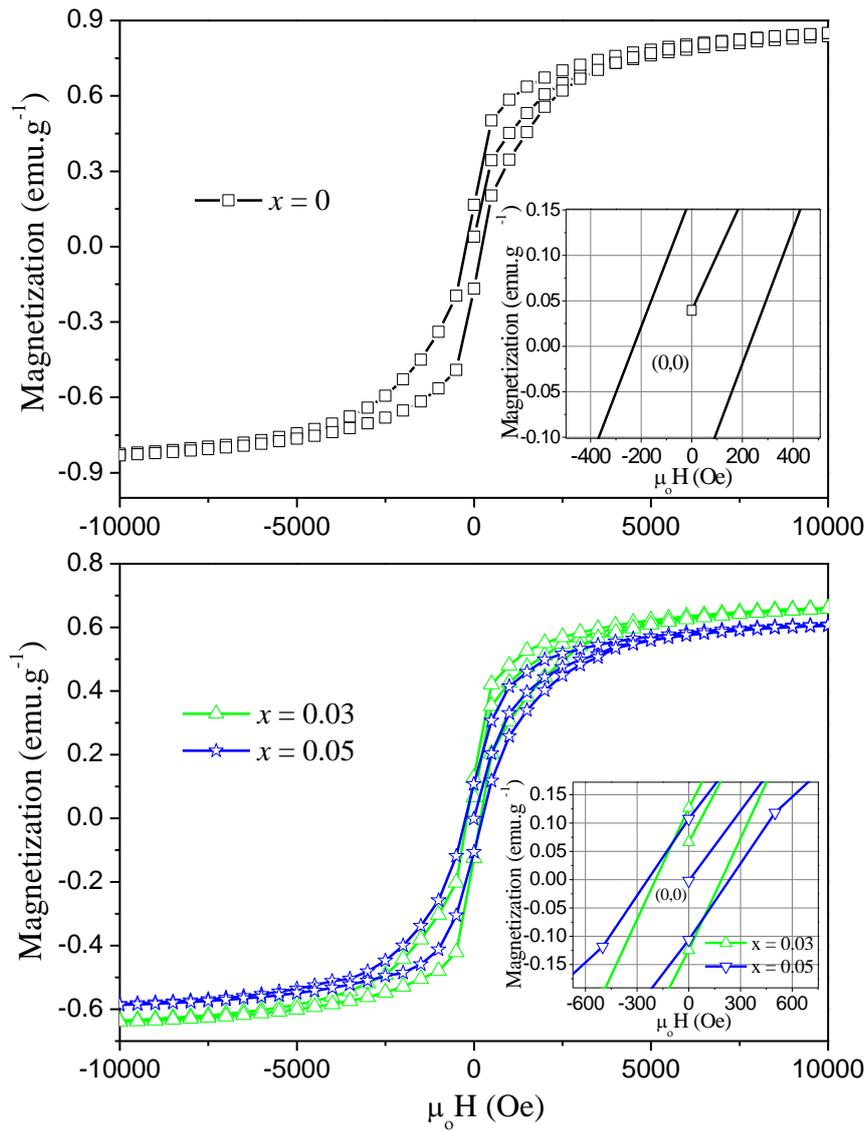

**Figure 8.** Magnetization as function of magnetic field of $Zn_{1-x}Cr_xO$ ($x$ = 0, 0.03 and 0.05) measured at room temperature. The inset is the enlarged $M(\mu_oH)$ curves at lower fields.

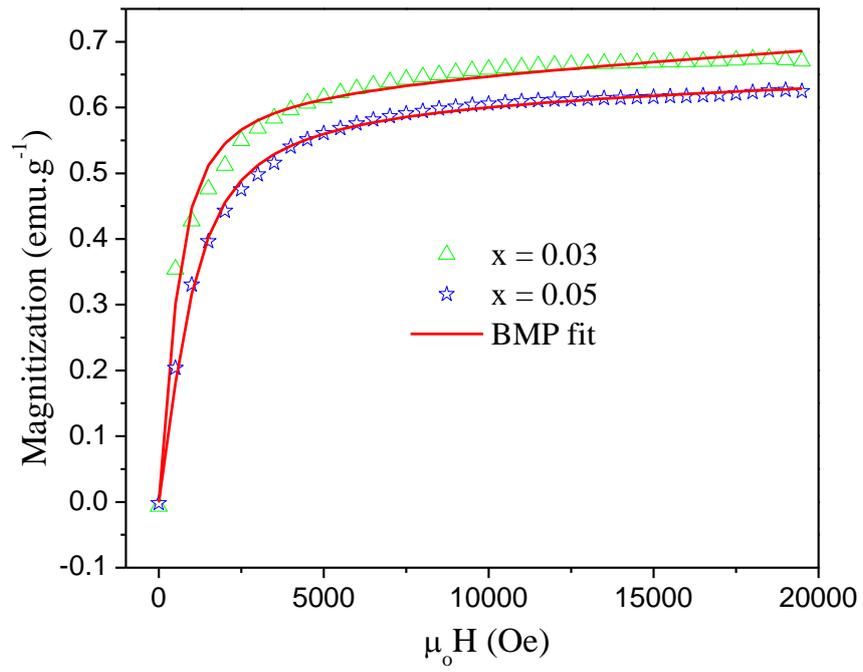

**Figure 9.** The symbols are the experimental data of $M(\mu_o H)$ curves of $Zn_{1-x}Cr_xO$ ($x = 0.03$ and 0.05) and red lines the fits to the experimental data using the BMP model.